\documentclass[10pt,letterpaper,amsmath,amsfonts,amssymb,aps,pra,reprint,superscriptaddress]{revtex4-1}
\usepackage[latin1]{inputenc}
\usepackage{bm}
\usepackage[pdftex]{graphicx}
\usepackage{color}
\usepackage{verbatim}
\usepackage{xargs}[2008/03/08]
\usepackage[colorinlistoftodos]{todonotes}
\usepackage{amsmath}
\usepackage{amsfonts}
\usepackage{amssymb}
\usepackage{bm}
\usepackage[pdftex]{graphicx}
\usepackage{etoolbox}
\newcommand{\op}[1]{\hat{\bm #1}}
\newcommand{\ket}[1]{\lvert #1\rangle}

\newcommand{\ipr}[2]{\langle #1 \vert #2 \rangle}

\global\long\def\i{{\rm i}}
\global\long\def\e{{\rm e}}
 \global\long\def\d{{\rm d}}
 \newcommandx\hd[1][usedefault, addprefix=\global, 1=]{\hat{d}_{#1}}
 \global\long\def\pd{\partial_{d}}
 \global\long\def\mi{\mathcal{I}}
 \newcommandx\np[1][usedefault, addprefix=\global, 1=]{N_{+}^{#1}}
 \newcommandx\nn[1][usedefault, addprefix=\global, 1=]{N_{-}^{#1}}
 \newcommandx\pp[1][usedefault, addprefix=\global, 1=]{p_{+}^{#1}}
 \newcommandx\pn[1][usedefault, addprefix=\global, 1=]{p_{-}^{#1}}
 \global\long\def\E#1{{\mathbb E}[#1]}
 \global\long\def\var#1{{\rm {Var}}[#1]}
 \global\long\def\cov#1{{\rm {Cov}}[#1]}
 \newcommandx\npm[1][usedefault, addprefix=\global, 1=]{N_{\pm}^{#1}}
 \newcommandx\ppm[1][usedefault, addprefix=\global, 1=]{p_{\pm}^{#1}}
 \global\long\def\t{{\rm tot}}
 \newcommandx\al[2][usedefault, addprefix=\global, 1=, 2=]{\alpha_{#1}^{#2}}
\newcommandx\ppp[1][usedefault, addprefix=\global, 1=]{p_{++}^{#1}}
\newcommandx\ppn[1][usedefault, addprefix=\global, 1=]{p_{+-}^{#1}}
\newcommandx\pnp[1][usedefault, addprefix=\global, 1=]{p_{-+}^{#1}}
\newcommandx\pnn[1][usedefault, addprefix=\global, 1=]{p_{--}^{#1}}
\global\long\def\npp{N_{++}}
\global\long\def\npn{N_{+-}}
\global\long\def\nnp{N_{-+}}
\global\long\def\nnn{N_{--}}
\global\long\def\sd{\psi_{d}}
\global\long\def\kd{|\sd\rangle}
\global\long\def\bas#1{|e_{#1}\rangle}
\newcommandx\pk[1][usedefault, addprefix=\global, 1=k]{p_{#1}}

\global\long\def\pd{\partial_{d}}
\global\long\def\iq{\mathcal{I}_{Q}}
\global\long\def\inn#1#2{\langle#1|#2\rangle}
\global\long\def\tk{\theta_{k}}
\global\long\def\ei#1{\e^{\i#1}}
\global\long\def\et{\ei{\tk}}
\global\long\def\sk{\sum_{k=1}^{n}}
\global\long\def\ic{\mathcal{I}_{C}}
\global\long\def\rk{\gamma_{k}}

\begin{document}
\title{Precision optical displacement measurements using biphotons}
\author{Kevin Lyons}
\author{Shengshi Pang}
\affiliation{Department of Physics and Astronomy, University of Rochester, Rochester, New York 14627, USA}
\affiliation{Center for Coherence and Quantum Optics, University of Rochester, Rochester, New York 14627, USA}
\author{Paul G. Kwiat}
\affiliation{Department of Physics, University of Illinois at Urbana-Champaign, Urbana, Illinois 61801-0380, USA}
\author{Andrew N. Jordan}
\affiliation{Department of Physics and Astronomy, University of Rochester, Rochester, New York 14627, USA}
\affiliation{Center for Coherence and Quantum Optics, University of Rochester, Rochester, New York 14627, USA}
\affiliation{Institute for Quantum Studies, Chapman University, 1 University Drive, Orange, CA 92866, USA}

\date{\today}

\begin{abstract}
We propose and examine the use of biphoton pairs, such as those created in parametric down conversion or four-wave mixing, to enhance the precision and the resolution of measuring optical displacements by position-sensitive detection.  
We show that the precision of measuring a small optical beam displacement with this method can be significantly enhanced by the correlation between the two  photons, given the same optical mode. 
The improvement is largest if the correlations between the photons are strong, and falls off as the biphoton correlation weakens.
More surprisingly, we find that the smallest resolvable parameter of a simple split detector scales as the inverse of the number of biphotons for small biphoton number (``Heisenberg scaling''), because the Fisher information diverges as the parameter to be estimated decreases in value.  One usually sees this scaling only for systems with many entangled degrees of freedom.
We discuss the transition for the split-detection scheme to the standard quantum limit scaling for imperfect correlations as the biphoton number is increased.  An analysis of an $N$-pixel detector is also given to investigate the benefit of using a higher resolution detector.   The physical limit of these metrology schemes is determined by the uncertainty in the birth zone of the biphoton in the nonlinear crystal.


\end{abstract}

\maketitle

\section{Introduction}
Measurements of the deflection or displacement of optical beams are useful in a wide range of experiments and applications; for example, optical beam deflections enable precision atomic force microscopy measurements with fairly modest experimental equipment \cite{Putnam1992}.  
In recent years, a variety of these methods have been developed using classical states of light.  
In particular, a number of schemes utilizing weak value amplification have been successful in measuring optical angular deflections as small as hundreds of femtoradians and linear displacements as small as tens of femtometers \cite{Dixon2009}, and have allowed for ultra-precise measurements to be made in relatively noisy environments \cite{Hosten2008}. 
For a review of weak value theory and experimental results see Ref.~\cite{Dressel2014}. 

It is well established that nonclassical states of light are capable of improving the precision of optical measurements (see, \emph{e.g.,}~\cite{Agarwal2013,giovannetti2006}).  
However, the vast majority of these quantum enhancements are in measurements having to do with phase or temporal properties of light, as opposed to spatial ones.  
An interesting series of both theory and experimental works \cite{Fabre2000, Treps2002, Treps2003, Barnett2003} have shown that squeezed states of light can be used to improve the sensitivity of split-detection displacement measurements by reducing the variance of the detected signal.  
This has been observed experimentally in both one dimensional \cite{Treps2002} and two dimensional \cite{Treps2003} displacements. 

In this paper we consider an alternative scheme for using quantum correlations to enhance the measurement of a spatial deflection or displacement.  
By using spatially entangled biphoton pairs, such as those created using parametric downconversion \cite{Schneeloch2015,Law2004,Fedorov2009}, we show it is possible to substantially reduce noise while using a simple experimental setup.
Namely, we find that correlations between photons in each pair allow for the average position of both photons to be determined more precisely than the individual positions of each photon.
With a judicious choice of measurement scheme, this allows for an enhancement in the determination of a small displacement parameter.
We note that fundamental measurement limits and the validity of Heisenberg's uncertainty relation using schemes with entangled probes and detectors were considered previously by Di Lorenzo \cite{DiLorenzo2013} and Bullock and Busch \cite{Bullock2014}.
They find that is possible to improve the sensitivity of a measurement using entanglement, which is in agreement with our analysis of a specific measurement scheme here.  
It is also interesting to consider that while continuous momentum and position correlations as quantum resources are relatively new in metrology, they have been at the center of our understanding of entanglement from a very early point in the development of quantum theory.
In particular, they were considered in the seminal work by Einstein, Podolsky, and Rosen \cite{Einstein1935}, and entangled pairs of the form we consider in this paper may properly be thought of as ``EPR pairs''.


The paper is organized as follows: 
in Sec.~\ref{sec:biphoton displacement} we discuss the entangled biphoton quantum state, and correlated position-position probability distribution which is used throughout the paper, along with a brief description of a possible experimental implementation.  
Generic enhancements over schemes using coherent states with the same mode are also discussed \footnote{When comparing entangled and uncorrelated photons in this work, we use the same transverse spatial mode for each case.  
	Namely, we do not consider differences in the minimum waist of a classical beam versus the minimum uncertainty of the average position of a SPDC biphoton pair for a given pump wavelength.  
	This allows us to quantify enhancements due to entanglement in a consistent way, and also avoids certain technical limitations such as saturation or burning of the detector for focused beams of light.}.
In Sec.~\ref{sub:split-detection} and Sec.~\ref{sub:npixel} we extend the treatment to cover the reasonable cases of split-detection and $N$-pixel detectors, respectively.
In particular, these sections demonstrate the enhancement over uncorrelated photons and the robustness of measurements to pixelation of the detection scheme.
Sec.~\ref{sub:scaling} discusses the case of very strong spatial correlation, and how this affects the scaling of the measurement resolution with the number of independent events.
This also provides insight into the limits of the measurement scheme.
In Sec.~\ref{sec:conclusion} we give our concluding remarks.

\section{Biphoton displacement measurement}\label{sec:biphoton displacement}

\begin{figure}
	\includegraphics[scale=.6]{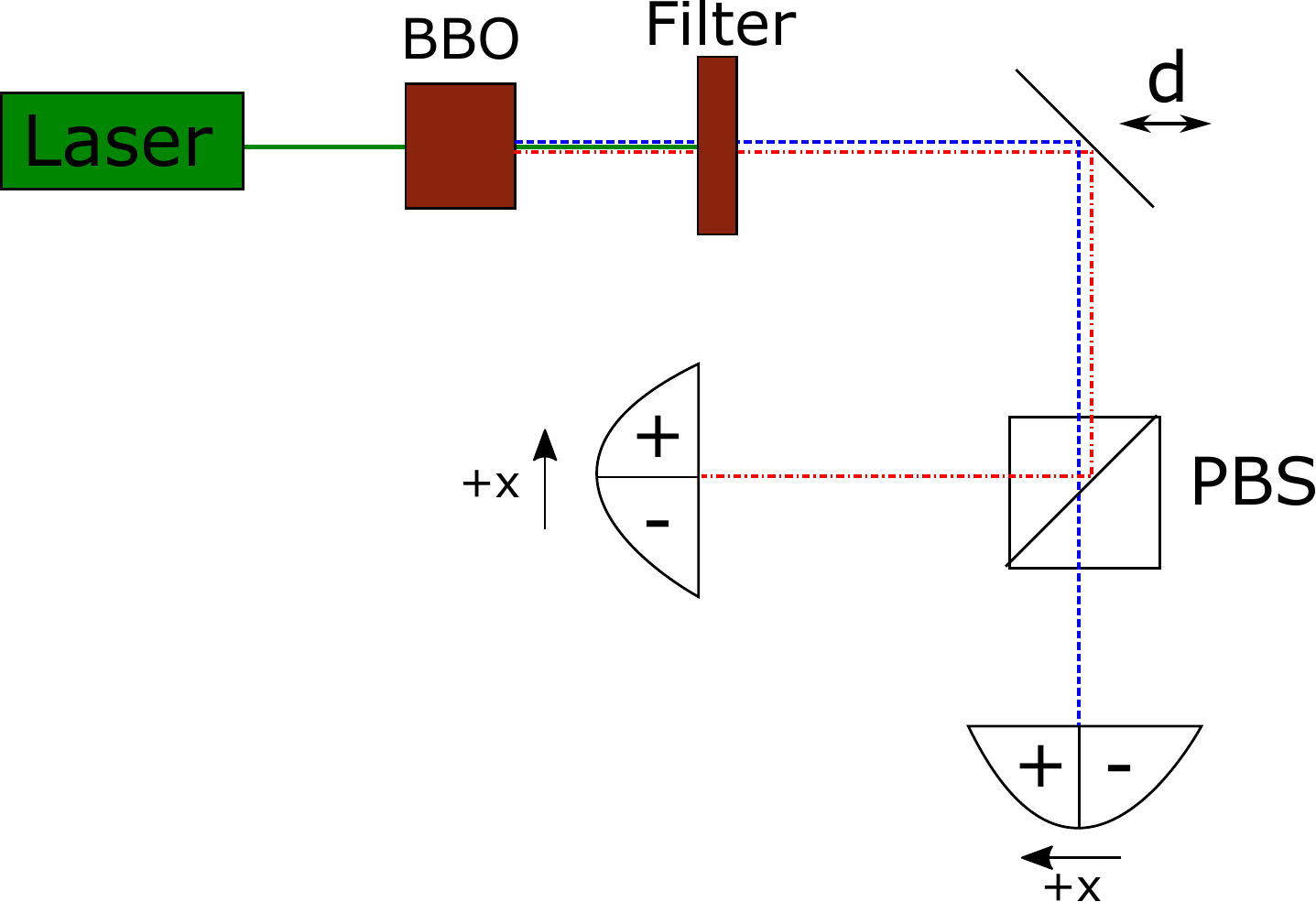}
	\caption{Schematic for a biphoton displacement measurement.  
		Photons from a laser source are converted into correlated pairs in a colinear type-II spontaneous parametric down conversion (SPDC) process (\emph{i.e.}, one in which two photons of orthogonal polarization are generated), for example with a BBO crystal.
      	Uncorrelated short wavelength photons (green solid line) are filtered out, while the horizontally polarized (red line) and vertically polarized (blue line) photons pass through to a movable mirror.  
		The shift $d$ of the mirror from the origin displaces the optical beam, and is the small, unknown parameter being measured with this apparatus.  
		Each photon in the pair is detected separately at a position-sensitive detector placed at the output ports of a polarizing beam splitter (PBS).
        By using coincidences between each detector it is possible to discard spurious events caused by lossy optics or imperfections in the PBS. 
		}
	\label{fig:setup}
\end{figure}

\begin{figure}
	\includegraphics[scale=1]{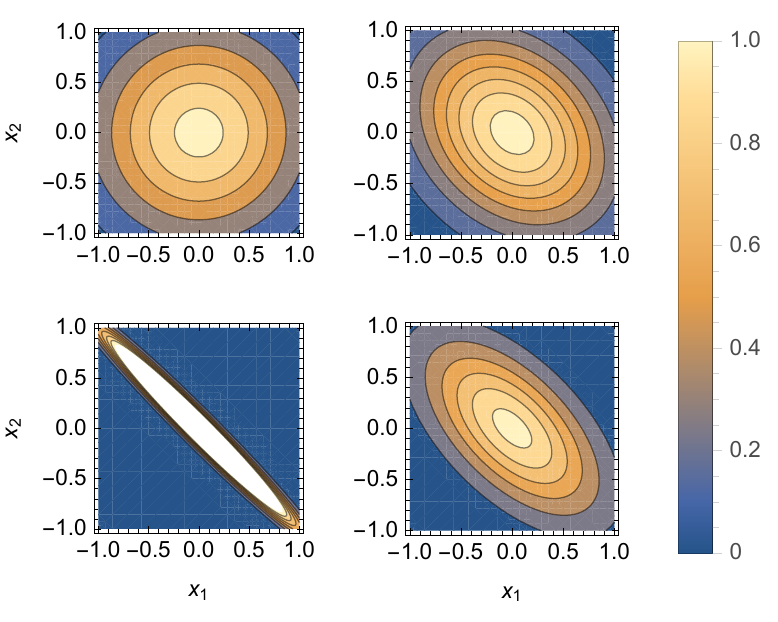}
	\caption{The probability distribution \eqref{eq:imperfectprofile} as a function of $x_1$ and $x_2$.  
		Clockwise from the upper left we have increasing correlation with $\epsilon/\sigma=1$, 0.75, 0.5, and 0.1.  
		As expected, $x_1$ and $x_2$ are increasingly anti-correlated with decreasing $\epsilon$. }
        \label{fig:distribution}
\end{figure}
As a simple, concrete model of a biphoton state exhibiting spatial correlations useful for displacement measurements, we consider the setup in Fig.~\ref{fig:setup}, which uses spontaneous parametric down conversion (SPDC) to generate the desired state.  
In SPDC, single ``pump'' photons are converted into two photons (typically referred to as the ``signal'' and ``idler'' modes), which are entangled due to constraints set by conservation of energy and conservation of momentum.  

The transverse state of biphotons created in SPDC can be approximated in the position basis as \cite{Schneeloch2015}
\begin{align}\label{eq:biphotonket}
\nonumber	\int  \text{d}x_1 \text{d}x_2 & \ket{x_1,x_2} \ipr{x_1,x_2}{\psi(d)} =  \\ 
\nonumber \frac{1}{\sqrt{\pi\sigma\epsilon}} \int \text{d}x_1 \text{d}x_2 & \exp\left(-\frac{(x_1 - x_2)^2}{4\sigma^2} \right) \\ 	
  \times \exp & \left(-\frac{(x_1 + x_2 - 2d)^2}{4\epsilon^2}\right) \op{a}^\dagger(x_1) \op{a}^\dagger(x_2)\ket{0,0},
\end{align}
where $d$ is the transverse displacement being measured, and $x_1$ and $x_2$ are the transverse position variables for each of the two photons, $\sigma$ is the pump beam waist, and $\epsilon$ is a parameter which describes the spatial correlation between the photons.
We note it is possible to switch $\epsilon$ and $\sigma$ in Eq.~\eqref{eq:biphotonket} by negating either $x_1$ or $x_2$, which is physically equivalent to a reflection of either the signal or idler mode.

The probability distribution follows simply from taking the norm squared of this state,
\begin{align}\label{eq:imperfectprofile}
\nonumber p({x_1,x_2}|d) &= \frac{1}{\pi\sigma\epsilon}\exp\left(\frac{-(x_{1} - x_{2})^2}{2\sigma^2}\right)\times \\
&\exp\left(\frac{-(x_{2} + x_{1} - 2d)^2}{2\epsilon^2}\right),
\end{align}
which is visualized in Fig.~\ref{fig:distribution}.
Note that an entangled pair of photons is necessary here to produce the correlated position distribution \eqref{eq:imperfectprofile}. If one uses some classical resources like coherent states, such a correlated distribution cannot be generated with this particular setup, since the PBS is a linear optical element and the photons at the two outputs are uncorrelated if the input photons are in a coherent state.
However, the results in this paper depend on this distribution only, rather than the specific optical physics or measurement scheme.
Since we will work in this fixed basis throughout this paper, the results of our analysis applies equally to any quantum or classical system which produces the distribution (\ref{eq:imperfectprofile}), \emph{i.e.,} for a judicious choice of measurement scheme it should be possible to replicate our results here using classically correlated beams.

For this profile, correlations set the constraint that if a photon is measured at position $x_1$, the probability distribution of the second photon is peaked at the position $x_2 = 2d - x_1$.  
In other words, the position of each photon is mirrored about the point $x = d$ up to a small uncertainty set by the parameter $\epsilon$. 
Hence, the smallness of $\epsilon$ determines the ``strength'' of the entanglement. 

In this paper we will use Fisher information \cite{Larsen2001} as the metric used to determine the overall measurement sensitivity for our biphoton distribution in various detector arrangements.  
From statistics, Fisher information is a measure of how sensitively the distribution depends on the parameter $d$.
The Fisher information per photon pair is simply calculated (for perfect detector resolution) as 
\begin{figure}
	\includegraphics[scale=1]{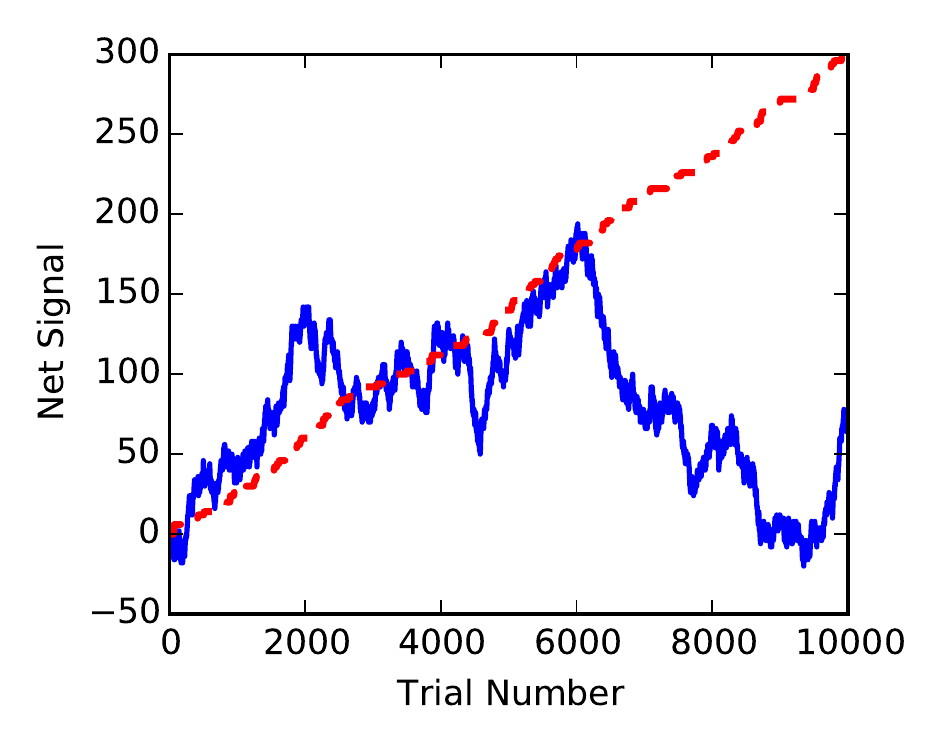}
	\caption{Net signal for biphotons after 10,000 independent events for split-detection.  
		The net signal here is analogous to the net steps from the origin in a random walk, with the probabilities of two steps backward, two steps forward, and zero steps per trial given by Eq.~\eqref{eq:biphotonprobs}.  
		The solid blue signal represents uncorrelated photon pairs, while the red dashed line represents entangled photons with $\epsilon/\sigma = 0.01$.}
	\label{fig:random walk}
\end{figure}
\begin{align}\label{eq:biphotonFisher}
	\nonumber	\mathcal{I}(d) &= \int^\infty_{-\infty}\int^\infty_{-\infty} \text{d}x_1 \text{d}x_2~ p({x_1,x_2}|d)\left( \partial_d \ln  p({x_1,x_2}|d)\right)^2, \\
	&= \frac{4}{\epsilon^2}
\end{align}
Note that the overall width of the input Gaussian profile $\sigma$ does not appear in this result.
Comparing with \eqref{eq:imperfectprofile} we see this amounts to a statement that the sensitivity of our measurement is not determined by the individual photon positions but rather the sum of their positions.  
This is in stark contrast to the case of uncorrelated photons, where only the mode profile width $\sigma$ determines the Fisher information, $\mathcal{I}_0(d) = 4/\sigma^2$.
The result \eqref{eq:biphotonFisher} can also be recovered using the quantum Fisher information given by \cite{Braunstein1996}
\begin{align}
	\mathcal{I}_Q(d) = 4\left(\ipr{\partial_d \psi}{\partial_d \psi} - \left|\ipr{\partial_d \psi}{\psi}\right|^2\right).
\end{align}
Since the state of biphoton \eqref{eq:biphotonket} has no $d$-dependent relative phases, it can be proven that the quantum Fisher information $I_Q(d)$ is equal to the classical Fisher information \eqref{eq:biphotonFisher} in this case (See Appendix B for the proof).

The Cram\`{e}r-Rao lower bound \cite{Larsen2001} defines a relationship between the Fisher information and the variance of a statistical estimator $\hat{d}$,
\begin{align}
\text{Var}(\hat{d}) \ge \frac{1}{\mathcal{I}(d)}.
\end{align}
Using the Cram\`{e}r-Rao lower bound to determine the minimum variance per biphoton pair and setting the minimum resolvable parameter $d_{min}$ equal to the square root of the variance, for $\nu$ independent measurements we have 
\begin{align}\label{eq:efficientdmin}
	d_{min} \sim \frac{\epsilon}{2\sqrt{\nu}},
\end{align}
for an efficient estimator (\emph{i.e.,}~an estimator which saturates the Cram\`{e}r-Rao lower bound).
Note that in this case each biphoton pair is an independent event, so there are $2\nu$ total photons in a given experimental run.
In the limit $\epsilon \rightarrow \sigma$, Eq.~\eqref{eq:imperfectprofile} is separable, and the distribution is identical to that of two uncorrelated photons.  
Hence, any advantage over the case of unentangled probes comes from a distribution with $\epsilon < \sigma$.
For a given input beam width $\sigma$, the enhancement in Fisher information due to entanglement is then
\begin{align}\label{eq:fisherratio}
	\frac{\mathcal{I}(d)}{\mathcal{I}_0(d)} = \frac{\sigma^2}{\epsilon^2}, 
\end{align}
where the comparison is between entangled biphotons and uncorrelated pairs of photons.  
For a comparison with single uncorrelated photons one can simply insert a factor of two due to the linearity of Fisher information (We note the uncorrelated position variance of a single photon is ${\rm Var}[x] = \sigma^2/2$ in this notation).
Sample split-detection signals highlighting the advantage of entanglement are shown in Fig.~\ref{fig:random walk}. 

The exact relation of $\sigma$ and $\epsilon$ can be further controlled by the placement of lenses before the polarizing beam splitter.  Quantum mechanics dictates that the small variance of $x_1 + x_2$ implies that the transverse wavenumber sum of the photons, $k_{x,1}+k_{x,2}$, has a large variance, since they obey an uncertainty principle, $\sigma_{x_+} \sigma_{k_+} \ge 1/2$ \cite{Schneeloch2015}.  The smallest uncertainty $\sigma$ of $x_1-x_2$ is determined by the pump beam width at the nonlinear crystal, and the smallest uncertainty $\epsilon$ of $x_1+x_2$ is given by $\epsilon_{\min} = \sqrt{9 w \lambda_p/10 \pi}$, where $w$ is the width of the crystal, and $\lambda_p$ is the wavelength of the pump beam \cite{Schneeloch2015}.

With the exception of results from expansions requiring $\epsilon/\sigma \ll 1$, one can simply replace $\epsilon$ with $\sigma$ in expressions throughout this work for generalization to uncorrelated pairs of photons.
Eq.~\eqref{eq:fisherratio} also provides a useful way to compare resources between the entangled and classical experiments.  Because Fisher information scales linearly with independent events, if we can achieve a given measurement precision with $\nu$ entangled photons, the classical equivalent will require $2 \nu (\sigma^2/\epsilon^2) $ independent photons to achieve the same precision.

The average position of the two photons $\hat{d} = (x_1 + x_2)/2$ is an efficient estimator for the parameter $d$, which is easily verified by direct calculation of the variance,
\begin{align}
	\mathbb{E}[\hat{d}] &= d, \\
	\mathbb{E}[\hat{d}^2] &= \frac{\epsilon^2}{4} + d^2.	
\end{align}
Hence, the variance per photon pair is $\epsilon^2/4$ and saturates the Cram\`{e}r-Rao lower bound.
Interestingly, if one could reduce the value of $\epsilon$ to arbitrarily low values, $d_{min}$ would become arbitrarily small for even a single biphoton pair. In the specific case of SPDC production, uncertainty in the birth zone of pairs leads to some minimum $\epsilon$.

We can also understand the effect of the detection scheme by considering the marginal probability distribution obtained by integrating over either $x_1$ or $x_2$ in Eq.~\eqref{eq:imperfectprofile},
\begin{align}
	p(x|d) = \sqrt{\frac{2}{\pi(\epsilon^2 + \sigma^2)}} \exp\left(\frac{-2(x-d)^2}{\epsilon^2 + \sigma^2} \right),
\end{align}
where subscripts have been dropped due to the symmetry of the distribution \eqref{eq:imperfectprofile}.
The Fisher information for the marginal distribution of each single photon if its twin is not measured is then 
\begin{align}\label{eq:margfisher}
	\mathcal{I}_\textrm{m}(d) = \frac{4}{\epsilon^2 + \sigma^2},
\end{align}
where the subscript m in $\mathcal{I}_\textrm{m}(d)$ indicates that it is the Fisher information of marginal distribution. 

Unlike Eq.~\eqref{eq:biphotonFisher}, the information is bounded in the limit $\epsilon \rightarrow 0$, and in the limit $\epsilon \rightarrow \sigma$ we recover the same result as for uncorrelated photons drawn from the full distribution \eqref{eq:imperfectprofile}.
Clearly the enhanced measurement sensitivity is not due only to the correlation between photons, but also the ability to detect the correlations for each event.

\subsection{Split-detection}\label{sub:split-detection}

The preceding results provide insight into the maximum achievable precision for a detector with perfect (\emph{i.e.,}~continuous) position resolution.  
Here we show that biphoton correlations can provide a benefit with a relatively simple split-detection scheme as well.  In a split-detection experiment one creates a detector out of two pixels and then uses the difference in counts between the two pixels to indicate the magnitude and direction of an optical beam shift.  A proposed setup is given in Fig.~1.  There, two split detectors are used together with a polarizing beam splitter (PBS), so that only events are counted where there is a coincidence detection event, so photons are simultaneously detected in both split-detectors as coincidences, and other (background) events are discounted.  For the presentation in the rest of the paper, we will discuss the results of a single split detector, which is theoretically the same, but not as technically easy to implement since both photons of a biphoton pair can both land on the same side of the detector.  To translate between the setups, both photons landing on the left $(-)$ or right $(+)$ side of the single-detector setup corresponds to the $++$ or $--$ two-detector events, whereas one photon landing left and the other right, corresponds to the two-detector $-+$ or $+-$ events.  The corresponding net signal is then the average of the split detector signals.  In Sec.~\ref{sub:npixel} we will extend this discussion to the more general case of a position-sensitive detector with $N$ pixels.

For the case of split-detection with a single detector, we introduce the probabilities $P(-2|d), P(0|d),$ and $P(2|d)$, which are the probabilities of two photons landing on the left half of the detector, one photon on each half of the detector, and two photons on the right half of the detector, respectively: 
\begin{align}\label{eq:biphotonprobs}
	\nonumber	P(-2|d) &= \int_{-\infty}^{0}\int_{-\infty}^{0}\text{d}x_1\text{d}x_2~p({x_1,x_2}|d), \\
	\nonumber	P(2|d) &= \int_{0}^{\infty}\int_{0}^{\infty}\text{d}x_1\text{d}x_2~p({x_1,x_2}|d),\\
	P(0|d) &= 1 - P(-2|d) - P(2|d).
\end{align} 
We ignore gaps between pixels throughout this work.
These integrals do not have closed form solutions, so we consider the limit $d \ll \epsilon$, where the probabilities can be expressed as   
\begin{align}
	\nonumber	P(\pm2|d) &\approx\frac{1}{4}+\frac{1}{2 \pi }\arctan\left(\frac{\epsilon }{2 \sigma }-\frac{\sigma }{2 \epsilon }\right)\pm d\sqrt{\frac{2}{\pi(\sigma ^2+\epsilon ^2)}}, \\
	P(0|d) &\approx\frac{1}{2}-\frac{1}{\pi}\arctan\left(\frac{\epsilon }{2 \sigma }-\frac{\sigma }{2 \epsilon }\right).
    \label{eq:splitProbs}
\end{align}
The resulting Fisher information for this discrete distribution is 
\begin{align}\label{eq:generalfisher}
	\mathcal{I}(d)\approx\frac{16}{\left(\epsilon ^2+\sigma ^2\right) \left(\pi+2 \arctan\left(\frac{\epsilon }{2 \sigma }-\frac{\sigma }{2 \epsilon }\right) \right)}.
\end{align}

Now, let us compute the correlation in the biphotons. A typical measure for the correlation between two variables $X$ and $Y$ is the correlation coefficient defined as
\begin{align}
\xi=\frac{\cov{X,Y}}{\sqrt{\var X\var Y}}.
\end{align}

From the joint distribution of a biphoton pair, Eq. \eqref{eq:imperfectprofile}, it can be derived that the correlation coefficient between the positions of two entangled photons is
\begin{align}\label{eq:genefisher}
	\xi=\frac{\epsilon ^2-\sigma ^2}{\epsilon ^2+\sigma ^2}.
\end{align}
This yields an explicit relation between the Fisher information $\mathcal{I}(d)$ and the correlation coefficient $\xi$:
\begin{align}\label{eq:Fisherinfosplit}
	\mathcal{I}(d)\approx\frac{16}{\left(\epsilon ^2+\sigma ^2\right) \left(\pi+2 \arcsin\xi \right)}.
\end{align}

By setting $\xi=0$ (\emph{i.e.}~, $\epsilon=\sigma$) in Eq. \eqref{eq:genefisher}, we find the Fisher information for uncorrelated photon pairs under split-detection,
\begin{align}\label{eq:splitUncorrelatedFisher}
	\mathcal{I}_0(d)\approx\frac{16}{\pi\left(\epsilon ^2+\sigma ^2\right)}=\frac{8}{\pi\sigma^2},
\end{align}
which is smaller than the perfect resolution case \eqref{eq:biphotonFisher} by a factor of $2/\pi$.
The increase (or decrease) in Fisher information due to entanglement is determined by the correlation coefficient:
\begin{align}\label{eq:relation}
	\frac{\mathcal{I}(d)}{\mathcal{I}_0(d)}\approx\frac{\pi}{\pi+2 \arcsin\xi }.
\end{align}

This equation characterizes the relation between the boost in Fisher information and the correlation of the entangled photons. It shows that when $\xi<0$, the Fisher information $\mathcal{I}(d)$ will be larger than that without correlation $\mathcal{I}_0(d)$. This implies an advantage of using biphotons with negative correlation (\emph{i.e.,}~spatially anticorrelated biphotons) in split-detection. It is widely known that correlation can enhance the precision of parameter estimation in the quantum metrology community. Eq. \eqref{eq:relation} can be perceived as a counterpart of that in the split-detection scheme. 

Meanwhile, Eq.~\eqref{eq:relation} also shows that when $\xi>0$, $\mathcal{I}(d)$ becomes smaller than $\mathcal{I}_0(d)$ instead. This contrast with the case of negative correlations implies that different types of correlations can have different effects on the Fisher information, and not all types of correlation in the biphotons are favorable to the performance of split-detection, even when the ``strength'' of the correlation is the same.

An important topic in quantum metrology is identifying and characterizing useful quantum resources for enhancing the sensitivity of measurements. The above analysis shows that the correlation coefficient between photons of a biphoton pair determines the gain in the Fisher information for detecting a beam displacement. This provides an explicit criterion for identifying useful correlation for split-detection with our setup.
 

In the limiting case $\xi\rightarrow-1$ (\emph{i.e.}, $\epsilon\rightarrow0$), the Fisher information can be extremely large, which is in accordance with Fig. \ref{fig:npixels}, while in the limiting case $\xi\rightarrow1$ (\emph{i.e.}, $\sigma\rightarrow0$), the Fisher information decreases to only half of that of uncorrelated photons. 
Note that in the limit $\epsilon\rightarrow0$, we need $d\rightarrow0$ as well, otherwise the assumption $d\ll\epsilon$ will be violated. 

We now investigate the maximum likelihood estimator for split-detection.
To begin, we introduce the binary random values $X_1$ and $X_2$ which can take on the value 0 or 1, and are mutually exclusive (\emph{i.e.,}~only one of them can take on the value 1 for a given event).  
$X_1$ represents the outcome where two photons land on the positive half of the detector and $X_2$ represents both landing on the negative half.
The relevant joint probabilities reduce to quantities we have already calculated above, namely $P(X_1 = 1, X_2 = 0|d) = P(2|d)$ and $P(X_1 = 0, X_2 = 1|d) = P(-2|d)$ so we are able to directly calculate the variance of the scaled split-detection estimator  
\begin{align}
	\hat{d} = \sqrt{\frac{\pi (\sigma^2 + \epsilon^2)}{8}} \left(X_1 - X_2\right).
\end{align}
Computing the variance yields precisely the inverse of Eq.~\eqref{eq:generalfisher} and so saturates the Cram\`{e}r-Rao lower bound. 
Hence, the split-detection estimator is efficient for measuring the displacement of biphoton pairs.

\subsection{$N$-pixel detector}\label{sub:npixel}
\begin{figure}
	\includegraphics[scale=.9]{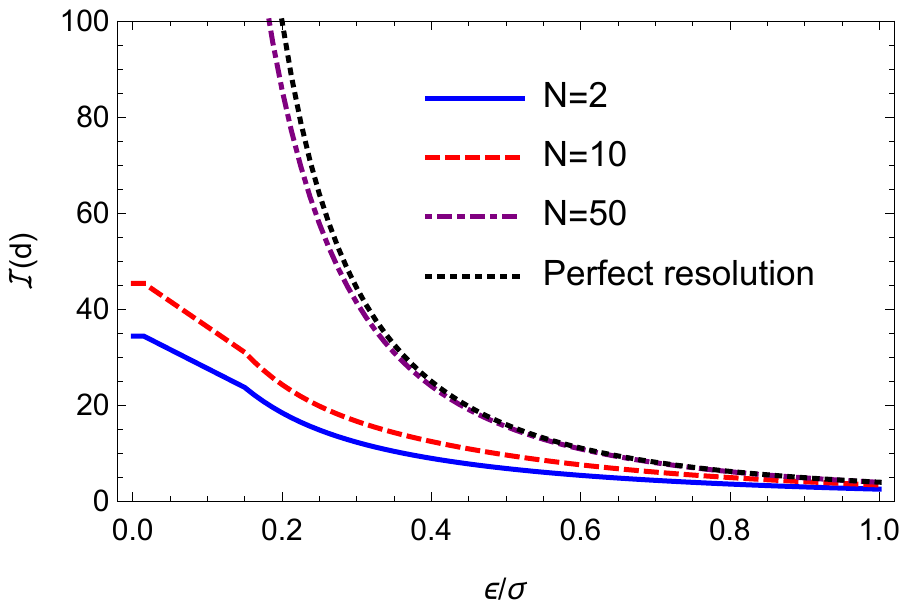}
	\caption{A comparison of the Fisher information (in arbitrary units) due to biphoton pairs incident on detectors with different numbers of pixels $N$, using the numerical value $\sigma=1$ and $d=0.05$.
	For each curve, the information for uncorrelated photons is given by the value of the curve for $\epsilon/\sigma = 1$.  
	The dotted black line represents perfect spatial resolution calculated by Eq.~\eqref{eq:biphotonFisher},  the solid blue curve represents split-detection ($N=2$), the red dashed curve represents a ten-pixel detector, and the case of fifty pixels is given by the purple dot-dashed curve.
	}\label{fig:npixels}
\end{figure}

For the case of a position-sensing detector with $N$ pixels, we may proceed similarly to the case of split-detection above.  
We will slightly change notation for convenience, with $P_{ij}$ being the probability of one photon each landing on pixels $i$ and pixel $j$ (with the case of $i=j$ being the case of both photons on the same pixel), and the dependence on the parameter $d$ is implied, as opposed to the explicit conditional probability notation used above.  
We note that similar calculations have been performed for the cases of weak value amplification and uncorrelated Gaussian beams \cite{Knee2014, Knee2015}.
If the pixels have a width $\Delta$, the probabilities may be calculated as 
\begin{align}
	\nonumber	P_{i,j\neq i} &= \int_{(i-1)\Delta}^{i\Delta}\text{d}x_1\int_{(j-1)\Delta}^{j\Delta} \text{d}x_2 p(x_1,x_2|d) \\
	&+ \int_{(i-1)\Delta}^{i\Delta}\text{d}x_2\int_{(j-1)\Delta}^{j\Delta} \text{d}x_1 p(x_1,x_2|d),
\end{align}
where the case of $i=j$ is equal to the first term alone.  
If we label the bins from $1$ to $N$, the joint probability distribution for a single event can be written as
\begin{align}
	p(X_{11},X_{12},...|d) = P_{11}^{X_{11}} P_{12}^{X_{12}}...,
\end{align}
where the random variables $X_{ij}$ are again binary with mutually exclusive outcomes for a single event.  
The indices $(i,j)$ are also subject to the constraint $i \le j$ to avoid double-counting outcomes.  
After some calculations analogous to the split-detection case, one finds the Fisher information per biphoton event is equal to
\begin{align}\label{eq:npixelFisher}
	\frac{\mathcal{I}(d)}{\nu} &= \sum_{i=1}^{N} \sum_{j\ge i}^{N} \frac{1}{P_{ij}}(\partial_d P_{ij})^2.
\end{align}
Results for the Fisher information due to our distribution incident on a ten-pixel and fifty-pixel detector compared to split-detection (two pixels) and to perfect spatial resolution are shown in Fig.~\ref{fig:npixels}. 
For these computations we use a detector width of $10\sigma$, and hence a pixel width of $10\sigma/N$, where $N$ is the number of pixels.
In general, it is possible to improve the resolution by using non-homogeneous pixel widths, but for simplicity we will not treat that case here. 

\subsection{Scaling with independent events}\label{sub:scaling}
From the above analysis (\emph{e.g.,}~Eqs.~\eqref{eq:biphotonFisher}-\eqref{eq:efficientdmin}), it is clear that for nonzero $\epsilon$ the minimum resolvable parameter scales as $\nu^{-1/2}$ for $\nu$ independent events (\emph{i.e.,} the usual shot-noise limit), with the enhancement appearing as a prefactor, as expected.  
However, taking the above treatment for the split-detection in the limit $\epsilon \rightarrow 0$ yields interesting results (keeping in mind that $\epsilon$ does have a minimum value it can take (see the discussion following Eq.~\eqref{eq:fisherratio})).  
For simplicity we rewrite Eq.~\eqref{eq:imperfectprofile} as 
\begin{align}
	p(x_1,x_2|d) = \sqrt{\frac{2}{\pi\sigma^2}}\exp\left(-\frac{(x_1 - x_2)^2}{2\sigma^2}\right)\delta(x_1 + x_2 - 2d).
\end{align}
The Fisher information for this probability distribution is infinite, implying that the statistical noise is completely suppressed and the estimation process becomes deterministic for the case of perfect spatial resolution, in agreement with Eq.~\eqref{eq:biphotonFisher}.
One can verify this by noting that the variance of the estimator $\hat{d} = (x_1 + x_2)/2$ is exactly zero. 
For the case of split-detection, we can exactly solve the integrals in Eq.~\eqref{eq:biphotonprobs}.  
Without loss of generality, we assume $d>0$;
\begin{align}\label{eq:perfectprobs}
\nonumber	P(2|d) &= \text{erf}\left(\frac{\sqrt{2} d}{\sigma}\right), \\
\nonumber	P(-2|d) &=0, \\
			P(0|d) &= 1 - P(2|d).
\end{align}
We note that these probabilities form a Bernoulli distribution (see, \emph{e.g.,}~\cite{Cahill2013, Larsen2001}) with $P(2|d)$ as the probability of a ``success'' and $1 - P(2|d)$ as the probability of a ``failure''.  
It follows immediately from the properties of a Bernoulli experiment with $\nu$ total trials for very small $d$:
\begin{align}\label{eq:perfectfisher}
	\mathcal{I}(d) \approx \sqrt[]{\frac{8}{\pi \sigma^2}} \cdot \frac{\nu}{d}.
\end{align}

Interestingly, the information increases as the parameter $d$ becomes smaller, and diverges as $d \rightarrow 0$.
This can be understood intuitively from Eq.~\eqref{eq:perfectprobs}, noting the outcome with one photon incident on each half of the detector is certain for the case of $d=0$.  
Therefore, events where two photons land on the same side of the detector give a great deal of information about the parameter's value being different from 0.

Eq.~\eqref{eq:perfectfisher} clarifies the boost in measurement precision of a small parameter due to extremely strong correlations.
Comparing the case of uncorrelated photons under split detection \eqref{eq:splitUncorrelatedFisher}, we note that the enhancement from entanglement can become arbitrarily large for an arbitrarily small parameter $d$. 

Another interesting enhancement to the measurement is the increase of the resolution. The resolution is the minimal resolvable parameter by the measurement with a given signal-to-noise ratio (SNR) $\mathcal{R}$. Since the Fisher information is the inverse of the minimum variance of the estimate, the resolution is related to the Fisher information via the SNR,
\begin{equation}\label{eq:snrdef}
\mathcal{R} \le \frac{d_{\min}}{\sqrt{\min\var{\hd}}}=d_{\min}\sqrt{\mi(d_{\min})},
\end{equation}
where the minimization before $\var{\hd}$ is over all possible unbiased estimators, and the second equality results from the definition of Fisher information.
A standard choice for the minimum SNR required to resolve a parameter is unity. Hence, by substituting Eq.~\eqref{eq:perfectfisher} into Eq.~\eqref{eq:snrdef}, the minimum resolvable $d$ turns out to be
\begin{align} \label{eq:resolution}
	d_{min} \approx \sqrt{\frac{\pi \sigma^2}{8}} \cdot \frac{1}{\nu}.
\end{align}
Eq. \eqref{eq:resolution} implies that the resolution reaches the Heisenberg-limited scaling when $\epsilon\rightarrow0$.

This is an interesting result. 
As is widely known, the standard quantum limit for the resolution scales as $\nu^{-\frac{1}{2}}$.
Typical enhancements over this scaling require large numbers of entangled quantum resources, \emph{e.g.}, squeezed states of light, in order for the resolution to reach the Heisenberg-limited scaling \cite{Barnett2003}. 
This contrasts with our setup, which only requires pairs of entangled photons, so there must be some substantially different mechanism from that of the other schemes for increasing the scaling of the resolution, as we now discuss.

It can be seen from the definition of SNR \eqref{eq:snrdef} that the key of improving the resolution is to increase the Fisher information. 
In most cases, the SNR is linear with the minimal resolvable parameter, so a large $N$ entangled photon state is used in order to increase the scaling of the Fisher information with respect to the number of photons, which results in an improved scaling of the measurement resolution. 
This is how the resolution in many optical metrology schemes is improved by using quantum resources like squeezed states.

However, in our protocol, when the parameter $d$ is decreased, the Fisher information increases. 
This changes the scaling of the Fisher information with respect to $d$, and similarly makes the SNR nonlinear with $d$. 
Hence, even though the scaling of the Fisher information with respect to the number of photons is unchanged, the resolution still can be significantly enhanced, and reach the Heisenberg-limited scaling as shown by Eq.~\eqref{eq:resolution}.

The above analysis also implies that the measurement precision, which is characterized by the minimum variance of the estimate, is not always equivalent to the resolution of the parameter to be measured, although they look quite similar and are sometimes used interchangeably to quantify the metrological performance. 
If the Fisher information is independent of the parameter, these two measures for the measurement are equivalent, for a given SNR; but if the Fisher information has dependence on the parameter, they can be rather different, as they are in the present case.
As our analysis of the biphoton displacement scheme for $\epsilon \rightarrow 0$ shows, the measurement precision is characterized by the Fisher information \eqref{eq:perfectfisher} which scales proportionally to $\nu$, as we would expect for a classical experiment.
However, the minimum resolvable parameter \eqref{eq:resolution} yields Heisenberg scaling proportional to $\nu^{-1}$.  
This is fundamentally due to the dependence of the Fisher information on the parameter $d$. 

\subsection{Small $\nu$ scaling}
\begin{figure}
\includegraphics[]{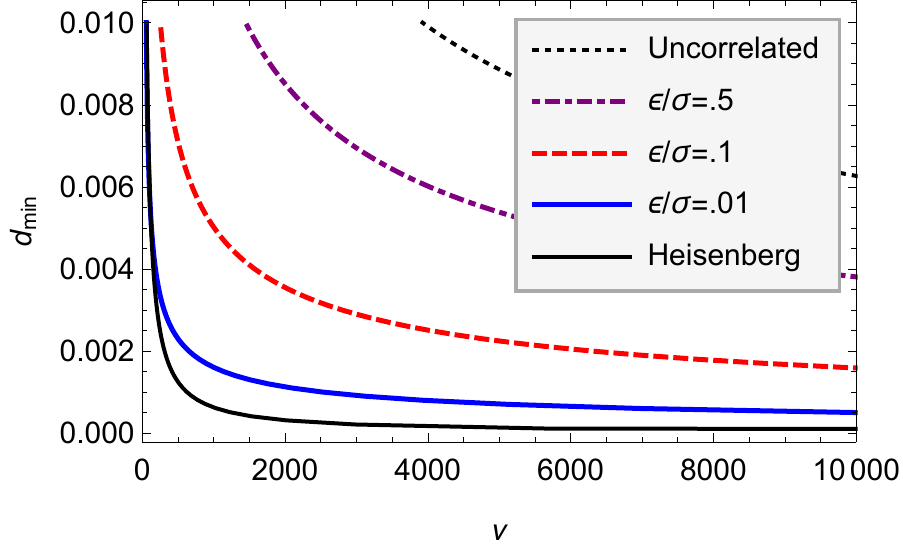}
\caption{The number of events needed to achieve a SNR of 1 under split-detection as a function of the parameter $d$ (in arbitrary units) for different values of $\epsilon/\sigma$.  From top to bottom the curves represent $\epsilon/\sigma = 1$ (\emph{i.e.,} uncorrelated), 0.5, 0.1, 0.05, and 0 (\emph{i.e.,} perfect correlations).  As expected from Eq.~\eqref{eq:heisenberg number}, Heisenberg scaling can only be observed for numbers of events much less than $\sigma/\epsilon$, e.g., the $\epsilon/\sigma = 0.01$ curve only matches the Heisenberg curve for $\nu<\sim 100$. }
\label{fig:crossover}
\end{figure}
In practice, because one cannot physically create a spatial profile which exactly has one photon hit a detector deterministically at one point conditional on the measurement of another entangled photon (\emph{i.e.,} $\epsilon=0$), the preceding analysis with delta function correlations is overly optimistic. 

As a simple model to consider more realistic imperfections, suppose the probabilities of ``success'' and ``failure'' may instead be expanded as
\begin{align}\label{eq:pracprob}
\nonumber P(\text{``success''}|d) &\approx \alpha + \beta d,\\
P(\text{``failure''}|d) &\approx 1-\alpha-\beta d,
\end{align}
with constant $\alpha$ and $\beta$ for some small but nonzero $\alpha$; $\alpha$ here is effectively a noise term indicating the probability of a success even when the parameter is zero, while $\beta$ determines the degree to which a change in the parameter correspondingly changes the probability of success. 
Comparing with Eq.~\eqref{eq:splitProbs}, we see we can determine $\alpha$ and $\beta$ simply by matching terms of the same order in $d$. 
The Fisher information of this probability distribution with respect to $d$ is
\begin{align}
\mi(d)\approx\frac{\beta^2\nu}{(\alpha + \beta d)(1-\alpha - \beta d)}.
\end{align}
Then the minimum resolvable parameter $d_{\min}$ can be derived from Eq.~\eqref{eq:snrdef} as
\begin{align}
d_{\min}=\frac{1-2\alpha+\sqrt{(1-2\alpha)^2+4\alpha(1-\alpha)\nu}}{2\beta\nu},
\end{align}
where $\nu \gg 1$ is assumed.
As in Sec.~\ref{sub:scaling}, we see the Fisher information is proportional to $\nu$ as we would expect for any experiment of $\nu$ independent events, but the dependence of the Fisher information on the parameter leads to interesting behavior in the scaling of the minimum resolvable parameter.

The scaling of $d_{\min}$ gradually transitions from $\nu^{-1}$ to $\nu^{-1/2}$ as $\alpha$ increases from zero. 
Therefore, the Heisenberg-limited scaling $\nu^{-1}$ is the limiting scaling of the resolution when $\alpha\ll1$. 
Hence, the condition to approximately achieve Heisenberg scaling is
\begin{align}
\alpha \ll \frac{1}{4\nu},
\end{align}
or equivalently for the case of $\epsilon \ll \sigma$,
\begin{align}\label{eq:heisenberg number}
\nu  \ll  \frac{\pi \sigma}{4 \epsilon}.
\end{align}
When $\nu$ increases to beyond this regime, the resolution of the measurement will scale as $\nu^{-1/2}$.
This has been confirmed numerically, shown in Fig.~\ref{fig:crossover}.
Note this result also implies that when we introduce any imperfection at all, the scaling will always be standard quantum limit in the asymptotic limit of large $\nu$ (but with a small prefactor).  This conclusions applies generically to added dephasing imperfections in Heisenberg scaling schemes \cite{bardhan2013effects,demkowicz2012elusive,jordan2015heisenberg}.

It is straightforward to show that an $N$-pixel detection scheme also reaches Heisenberg scaling in the limit of very strong entanglement.  
In the same notation as Sec.~\ref{sub:npixel}, we expand the probabilities to first order in $d$,
\begin{align}
P_{ij} \approx \alpha_{ij} + \beta_{ij} d.
\end{align}
Inserting these probabilities into Eq.~\eqref{eq:npixelFisher}, we see immediately that if even a single term in the sum satisfies $a_{ij} = 0$, the Fisher information in the limit of very small $d$ scales as $d^{-1}$ and hence the minimum resolvable parameter scales as $\nu^{-1}$ for small $\nu$.

We stress that if one has higher resolution, as can be achieved with more pixels, then the measurement precision can be increased.  Since the perfect detector limit of the entangled biphoton case scales as the standard quantum limit (\ref{eq:efficientdmin}) but with a prefactor of $\epsilon$, we have the following ordering of precision from greatest to least given a fixed number of photons: (1) Standard quantum limit scaling of the biphoton case with a perfect resolution detector, (2) Heisenberg scaling of the biphoton case with a split (2-pixel) detector, (3)  Standard quantum limit scaling of independent photons with a perfect resolution detector, (4)  Standard quantum limit scaling of independent photons with a split (biphoton) resolution detector.  This indicates that the scaling behavior is a secondary consideration to the Fisher information.

\section{Discussion and Conclusion}\label{sec:conclusion}
By using spatially entangled biphoton pairs, it is possible to reduce the quantum noise intrinsic to optical metrology schemes.  
While this is true in the case of split-detection, which is a typical detector setup for optical displacement measurements, we have shown that an even larger benefit can be obtained as one improves spatial resolution.   The ultimate physical limitation for this method is the position uncertainty in the birth zone of the biphoton.  We have also seen that it is possible to change the scaling in precision for split detection measurement between standard quantum limit and Heisenberg if one uses entangled biphotons instead of independent photons for the same mode profile, for relatively small photon number.  
This advantage comes from the fact that the sum of the biphoton positions is determined more precisely than for uncorrelated photons.

The detection and estimation scheme used in the paper is to count the number of times that the biphoton arrive at the left side, right side or at different sides of the detector and then extract the information of the beam displacement from the counts. We proposed an experiment using coincidence counting and two split detectors to implement this theory. This method is statistically optimal for this experimental setup.  A possible alternative approach is to consider the distribution of each single photon, estimate the displacement from it, and then average the estimates from the two photons. This approach also utilizes the correlation of the entangled photons, since the two estimates are correlated by the photon correlation.  This second approach is analyzed in detail in Appendix A, and we show the estimation precision can also reach the Cram\'er-Rao bound, and therefore matches the precision of split-detection detailed in the main text.  

\section{Acknowledgements}
This work was supported by Army Research Office Grant No. W911NF-13-1-0402.
We thank John Howell, Justin Dressel, James Schneeloch, and Jeff Tollaksen for discussions.

\bibliography{biphotons}
\pagebreak\setcounter{equation}{0} \renewcommand{\theequation}{S\arabic{equation}} \onecolumngrid \setcounter{enumiv}{0}

\section*{Appendix A: Estimation from marginal distributions}

In this Appendix, we study an alternative approach for estimating the  displacement of biphoton beam using split-detection. We consider the maximum likelihood estimation for the marginal distribution of each photon in a biphoton pair under split-detection and average the two estimates. We show here this procedure can also attain the Cram\'er-Rao bound when the estimates of the marginal distributions are evenly averaged.

According to Eq. \eqref{eq:imperfectprofile}, the marginal distribution of either photon in a biphoton pair is
\begin{equation}
p(x|d)=\sqrt{\frac{2}{\pi(\sigma^{2}+\epsilon^{2})}}\exp\left(-\frac{2(d-x)^{2}}{\sigma^{2}+\epsilon^{2}}\right),\label{eq:marginal distribution}
\end{equation}
where the subscripts $1,\,2$ are dropped since the marginal distributions
for the two photons are the same due to the symmetry of $p(x_{1},x_{2}|d)$.

For split-detection, the probability to find either photon of a biphoton pair at the
left or right side of the detector is, respectively,
\begin{equation}
\begin{aligned}p_{-} & =\int_{-\infty}^{0}p(x|d)\d x=\frac{1}{2}\left(1-\text{erf}\left(\frac{\sqrt{2}d}{\sqrt{\sigma^{2}+\epsilon^{2}}}\right)\right) \approx\frac{1}{2}-\sqrt{\frac{2}{\pi(\sigma^{2}+\epsilon^{2})}}d,\\
p_{+} & =\int_{0}^{+\infty}p(x|d)\d x=\frac{1}{2}\left(1+\text{erf}\left(\frac{\sqrt{2}d}{\sqrt{\sigma^{2}+\epsilon^{2}}}\right)\right)\approx\frac{1}{2}+\sqrt{\frac{2}{\pi(\sigma^{2}+\epsilon^{2})}}d,
\end{aligned}
\label{eq:2}
\end{equation}
where the approximation assumes $d\ll\sqrt{\sigma^{2}+\epsilon^{2}}$.

The Fisher information for the marginal distribution of either photon
is
\begin{equation}
\mi_\mathrm{m}=\frac{8}{\pi\left(\sigma^{2}+\epsilon^{2}\right)}.\label{eq:Fisher information}
\end{equation}
The subscript m in $\mathcal{I}_\textrm{m}(d)$ indicates it is the Fisher information of marginal distribution.

For the maximum likelihood estimation (MLE),
\begin{equation}
\nn\pd\log\pn+\np\pd\log\pp=0,
\end{equation}
where $\nn$ and $\np$ are the numbers of the photons that arrive at the left or right side of the detector, respectively, so the estimator for the marginal distribution is
\begin{equation}
\hd=\sqrt{\frac{\pi(\sigma^{2}+\epsilon^{2})}{8}}\frac{\np-\nn}{\np+\nn}.
\end{equation}

When there are $N$ pairs of biphotons in total, $N=\np+\nn$,
and the distribution of $\np,\,\nn$ is
\begin{equation}
p(\np,\nn)=\binom{N}{\np}\pp[\np]\pn[\nn].
\end{equation}
The variances of $\np,\,\nn$ and the covariance between them can be obtained:
\begin{equation}
\begin{aligned}\var{\npm} & =N\pp\pn,\\
\cov{\np,\nn} & =-N\pp\pn,
\end{aligned}
\end{equation}
so the variance of $\hd$ is
\begin{equation}
\begin{aligned}\var{\hd} & =\frac{\pi(\sigma^{2}+\epsilon^{2})}{8N^{2}}(\var{\np}+\var{\nn}-2\cov{\np,\nn})\\
 & \approx\frac{\pi(\sigma^{2}+\epsilon^{2})}{8N}.
\end{aligned}
\end{equation}
The inverse of $\var{\hd}$ is exactly the Fisher information \eqref{eq:Fisher information}
(scaled by $N$), so it verifies that the estimator $\hd$ reaches
the Cram\'er-Rao bound.

Since we have two photons from a biphoton pair, we can do some
average of the estimates from these two photons. We denote the estimators
of the two photons are $\hd[1]$ and $\hd[2]$ respectively, each
satisfying the distribution \eqref{eq:marginal distribution}. If we average them with some weights $\al[1]$ and $\al[2]$,
then the total estimator is
\begin{equation}
\hd[\t]=\al[1]\hd[1]+\al[2]\hd[2],\;\al[1]+\al[2]=1.
\end{equation}
The variance of this total estimator is
\begin{equation}
\begin{aligned}\var{\hd[\t]} & =\al[1][2]\var{\hd[1]}+\al[2][2]\var{\al[2]}\\
 & +2\al[1]\al[2]\cov{\hd[1],\hd[2]}.
\end{aligned}
\end{equation}
We have obtained the $\var{\hd[1,2]}$ above, and the covariance $\cov{\hd[1],\hd[2]}$
is
\begin{equation}
\begin{aligned}\cov{\hd[1],\hd[2]} & =\E{\hd[1]\hd[2]}-\E{\hd[1]}\E{\hd[2]}\\
 & =\E{\hd[1]\hd[2]}-d^{2},
\end{aligned}
\label{eq:3}
\end{equation}
where we have used the fact $\E{\hd[1]}=\E{\hd[2]}=d$. From here we must calculate $\E{\hd[1]\hd[2]}$.

From the joint spatial distribution \eqref{eq:imperfectprofile},
we can work out the joint distribution for the split-detection results,
which can be denoted as $\ppp,\,\ppn,\,\pnp,\,\pnn$. Then $\pnn=P(-2|d),\,\ppp=P(2|d),$ and $\ppn+\pnp=P(0|d)$, which were defined in Eq. \eqref{eq:biphotonprobs}, and
\begin{equation}
\begin{aligned}\pp & =\ppp+\ppn=\ppp+\pnp,\\
\pn & =\pnn+\pnp=\pnn+\ppn,
\end{aligned}
\label{eq:1}
\end{equation}
according to the definition of marginal distribution and the symmetry
of the biphoton distribution.

$\E{\hd[1]\hd[2]}$ can be expanded as
\begin{equation}
\begin{aligned}\E{\hd[1]\hd[2]} & =\frac{\pi(\sigma^{2}+\epsilon^{2})}{8N^{2}}\E{(\np[(1)]-\nn[(1)])(\np[(2)]-\nn[(2)])}\\
 & =\frac{\pi(\sigma^{2}+\epsilon^{2})}{8N^{2}}{\rm E}[(\npp+\npn-\nnp-\nnn)\\
 & \times(\npp+\nnp-\npn-\nnn)],
\end{aligned}
\end{equation}
where $\npp$ is the number of times that both photons of a biphoton pair hit the right side of the detector, and similarly for $\npn,\,\nnp,\,\nnn$.
The joint probability distribution for $\npp$, $\npn$, $\nnp$,
$\nnn$ is a multinomial distribution
\begin{equation}
\begin{aligned} & p(\npp,\npn,\nnp,\nnn)\\
 & =\frac{N!}{\npp!\npn!\nnp!\nnn!}\ppp[\npp]\ppn[\npn]\pnp[\nnp]\pnn[\nnn],
\end{aligned}
\end{equation}
so,
\begin{equation}
\begin{aligned}\E{N_{ij}} & =Np_{ij},\\
\E{N_{ij}N_{i^{\prime}j^{\prime}}} & =N(N-1)p_{ij}p_{i^{\prime}j^{\prime}}+Np_{ij}\delta_{ij,i^{\prime}j^{\prime}},
\end{aligned}
\end{equation}
where $i,j,i^{\prime},j^{\prime}=+\,{\rm or}\,-$. Therefore,
\begin{equation}
\begin{aligned}
 \E{\hd[1]\hd[2]}& =\frac{\pi(\sigma^{2}+\epsilon^{2})}{8N^{2}}[N(N-1)(\ppp+\ppn-\pnp-\pnn)\\
 & \times(\ppp+\pnp-\ppn-\pnn)+N(\ppp-\ppn-\pnp+\pnn)]\\
 & =\frac{\pi(\sigma^{2}+\epsilon^{2})}{8N}[(N-1)(\pp-\pn)^{2}+2(\ppp+\pnn)-1],
\end{aligned}
\label{eq:4}
\end{equation}
where Eq. \eqref{eq:1} has been used at the second equality.

The $(\pp-\pn)^{2}$ term can be dropped in Eq. \eqref{eq:4} because
it is proportional to $d^{2}$ according to \eqref{eq:2}, and the
$-d^{2}$ term can also be dropped in Eq. \eqref{eq:3} for the
same reason, so,
\begin{equation}
\cov{\hd[1],\hd[2]}\approx\frac{\pi(\sigma^{2}+\epsilon^{2})}{8N}[2(\ppp+\pnn)-1].
\end{equation}
It can be derived by direct calculation that
\begin{equation}
\begin{aligned}\ppp & =\frac{1}{4}+\frac{\arcsin\xi}{2\pi}+d\sqrt{\frac{2}{\pi(\sigma^{2}+\epsilon^{2})}},\\
\pnn & =\frac{1}{4}+\frac{\arcsin\xi}{2\pi}-d\sqrt{\frac{2}{\pi(\sigma^{2}+\epsilon^{2})}}.
\end{aligned}
\end{equation}
Thus,
\begin{equation}
\cov{\hd[1],\hd[2]}=\frac{(\sigma^{2}+\epsilon^{2})\arcsin\xi}{4N}.
\end{equation}

The result of $\var{\hd[\t]}$ is therefore,
\begin{equation}
\begin{aligned}\var{\hd[\t]} & =\frac{\pi(\sigma^{2}+\epsilon^{2})}{8N}(\al[1][2]+\al[2][2]+\frac{4}{\pi}\al[1]\al[2]\arcsin\xi)\\
 & =\frac{\pi(\sigma^{2}+\epsilon^{2})}{8N}[1-2\al[1]\al[2](1-\frac{2}{\pi}\arcsin\xi)],
\end{aligned}
\end{equation}
where we have used $\al[1]+\al[2]=1$.

What is the minimum value of $\var{\hd[\t]}$? Since $\frac{2}{\pi}\arcsin\xi-1\leq0$,
$\var{\hd[\t]}$ is minimized when $\al[1]\al[2]$ is maximized. Using
$\al[1]+\al[2]=1$, we have $\al[1]\al[2]\leq\frac{1}{4}$. The equality
holds when $\al[1]=\al[2]=\frac{1}{2}$. So, the minimum variance
of $\hd[\t]$ is
\begin{equation}
\var{\hd[\t]}_{\min}\approx\frac{\pi(\sigma^{2}+\epsilon^{2})}{8N}\Big(\frac{1}{2}+\frac{1}{\pi}\arcsin\xi\Big).
\end{equation}

This coincides with the Fisher information $\mi(d)$ in Eq. \eqref{eq:Fisherinfosplit} of
the main manuscript. It implies that the optimal precision of split
detection of biphotons can be attained via estimation from the two
marginal distributions followed by properly averaging the two estimates.

\section*{Appendix B: Fisher information: quantum vs. classical}

In this Appendix, we study when the quantum Fisher information of
a parameter-dependent state can be achieved by projective measurements.
Suppose the state of interest is $\kd$. If we want to estimate $d$
by performing a projective measurement on this state, along some $d$-independent
basis $\{\bas 1,\cdots,\bas n\}$, where $n$ is the dimension of
the system, we can expand the state $\kd$ along that basis as
\begin{equation}
\kd=\sk\sqrt{\pk}\et\bas k,\label{eq:state sd}
\end{equation}
where $\pk,\,\tk,\,\,k=1,\cdots,n$ depend on $d$. Then the probability
of obtaining the $k$-th result is $\pk$, and the classical Fisher
information of estimating $d$ by such a projective measurement is
\begin{equation}
\ic=\sk\frac{(\pd\pk)^{2}}{\pk}.\label{eq:cfi}
\end{equation}

The quantum Fisher information of $\kd$ is the maximum Fisher information
of estimating $d$ over all possible POVMs (not only projective measurements)
on $\kd$, and it can be obtained as
\begin{equation}
\iq=4(\inn{\pd\sd}{\pd\sd}-|\inn{\sd}{\pd\sd}|^{2}).\label{eq:def-qfi}
\end{equation}
Substituting Eq. \eqref{eq:state sd} into \eqref{eq:def-qfi}, the quantum
Fisher information of $\kd$ is
\begin{equation}
\begin{aligned}\iq & =4\Big(\sk\Big|\frac{\pd\pk}{2\sqrt{\pk}}+\i\sqrt{\pk}\pd\tk\Big|^{2}-\Big|\sk\sqrt{\pk}\Big(\frac{\pd\pk}{2\sqrt{\pk}}+\i\sqrt{\pk}\pd\tk\Big)\Big|^{2}\Big)\\
 & =\sk\frac{(\pd\pk)^{2}}{\pk}+4\Big[\sk\pk(\pd\tk)^{2}-\Big|\sk\pk\pd\tk\Big|^{2}\Big].
\end{aligned}
\end{equation}
Hence,
\begin{equation}
\iq-\ic=4\Big[\sk\pk(\pd\tk)^{2}-\Big|\sk\pk\pd\tk\Big|^{2}\Big].
\end{equation}
Applying the Cauchy-Schwartz inequality,
\begin{equation}
\Big|\sk\pk\pd\tk\Big|^{2}\leq\sk\pk(\pd\tk)^{2};
\end{equation}
the equality holds only when
\begin{equation}
c\sqrt{\pk}=\sqrt{\pk}\pd\tk,\,\forall k,
\end{equation}
where $c$ is a constant. So, when $\iq=\ic$, the solution to $\tk$ is $\tk=cd+\rk$,
where $\rk$ is a constant independent of $d$ for each $k$. In this case,
\begin{equation}
\et=\ei{cd}\ei{\rk}.
\end{equation}
Note that $\ei{cd}$ is a global phase of $\kd$ which can be dropped,
and $\ei{\rk}$ is a phase independent of $d$. Therefore, the quantum
Fisher information is equal to the classical Fisher information only when the state has no parameter-dependent
relative phases in the basis of the measurement.
\end{document}